\documentclass{osa-article}

\journal{osajournal}

\articletype{Research Article}

\usepackage{caption}
\usepackage{subcaption}
\usepackage{comment}
\newcommand{\dg}[0]{^\circ}

\usepackage{graphicx}

\usepackage[export]{adjustbox}

\usepackage{multirow}
\usepackage{graphicx}
\usepackage{tikz}
\usepackage{placeins}
\usepackage{amsmath}
\usepackage[utf8]{inputenc}
\usepackage{here}

\usepackage[stable]{footmisc}

\usepackage{upgreek}

\begin{document}

\title{Electro-optic sampling based characterization of broad-band high efficiency THz-FEL}

\author{M. Lenz,\authormark{1,2}, A. Fisher\authormark{1}, A. Ody\authormark{1}, Y. Park\authormark{1}, and P. Musumeci \authormark{1,3}}

\address{\authormark{1}Department of Physics and Astronomy, University of California at Los Angeles, 90095-1547

\authormark{2}mlenz@g.ucla.edu
\authormark{3}musumeci@physics.ucla.edu}


\begin{abstract}
Extremely high beam-to-radiation energy conversion efficiencies can be obtained in a THz FEL using a strongly tapered helical undulator at the zero-slippage resonant condition, where a circular waveguide is used to match the radiation group velocity to the electron beam longitudinal velocity. In this paper we report on the first electro-optic sampling (EOS) based measurements of the broadband THz FEL radiation pulses emitted in this regime. The THz field waveforms are reconstructed in the spatial and temporal domains using multi-shot and single-shot EOS schemes respectively. The measurements are performed varying the input electron beam energy in the undulator providing insights on the complex dynamics in a waveguide FEL. 
\end{abstract}


\section{Introduction} \label{sec:Introduction}


In the 0.1-10 THz range of the electromagnetic spectrum, high power solid-state sources are relatively scarce and free-electron lasers (FELs) are very attractive due to their tunability, high peak and average power and because electron beam and undulator parameters are readily achievable with current technology \cite{mullerschwarz,galleranobiedron}.

At these long wavelengths, FELs are inherently limited by the slippage effect in which the radiation moves forward with respect to the particles by one radiation wavelength each undulator period \cite{eliasUCSB,ramianUCSB,oepts_FELIX,gallerano_CFEL,jeong_KAERI,NovosibirskTHz, THzElbe,CTFEL,FELICHEM,polfel,nause20216,PITzFEL,SeededTHzFEL}. In order to counteract this, relatively long (many wavelength) electron beams are used to ensure that radiation amplification can continue throughout the entire undulator length. However longer electron beams imply lower peak currents and therefore smaller gain and ultimately lower output power. Recently it has been shown that by using a waveguide to match the e-beam and the radiation group velocity, it is possible to maintain temporal overlap between the electromagnetic pulse and a short electron bunch in the undulator so that enhanced energy exchange might be achieved \cite{curry:meterscale, snively2019broadband}. The waveguide has the additional advantage of eliminating radiation diffraction which can also severely impact the coupling between e-beam and radiation at millimeter wavelengths. Strong undulator tapering can then be used to maintain the resonant condition and efficiently convert a large fraction of the particles' kinetic energy to coherent radiation \cite{duris:TESSA}. An experiment in this regime at the UCLA Pegasus photoinjector beamline showed up to 10\,\% extraction efficiency \cite{TESSATRON}. The initial measurements included characterization of the emitted pulse using a pyrometer to retrieve the THz energy and a multi-shot interferometer to record the spectral content of the radiation. 

The data confirmed how the THz-FEL at the zero-slippage condition is characterized by very large (50\,\%) relative bandwidths in contrast to the narrower-band (1\,\%) traditional FEL schemes supporting the amplification of shorter pulses. Just as in other cases where coherent broadband radiation is involved (for example the laser-based THz sources \cite{hebling}), it is important to fully characterize the pulse in time-domain and retrieve both amplitude and phase of the THz waveform. The field amplitude controls the pulse peak intensity and is ultimately responsible for the non linearities in material response. Measurement of the field phase relays information about pulse chirp and about the transport dispersion. In our particular case, the phase also carries details of the emission process, as the FEL amplification is seeded using  compressed electron bunches and we therefore expect the THz pulse to be phase-locked with the electrons. 


In this paper we describe the application of the electro-optic sampling (EOS) technique to the characterization of the radiation pulses from the UCLA Pegasus THz-FEL \cite{TESSATRON}. EOS has been successfully used in many accelerator beamlines to characterize radiation from electron beams (transition radiation, diffraction radiation and other coupling schemes \cite{hoffmann:CTR,PhysRevLett.123.094801, brussaard:SPP}) and in laser-based THz generation setups \cite{nahata,hirori2011single}. To our knowledge this is the first time that a THz-FEL operating in the zero-slippage regime is characterized by EOS. We used a multishot scan and a single shot spectral encoding scheme to characterize the THz FEL waveforms spatially and temporally respectively. Measurements are conducted as a function of the input beam energy, shedding light on the FEL dynamics in this regime. As the energy is increased from the zero-slippage condition, the data reveals clear pulse splitting with an earlier higher frequency pulse followed by a lower frequency pulse. This behavior, which could not be distinguished simply using an autocorrelation trace, is explained as due to the two intersections of the e-beam phase velocity with the quadratic waveguide dispersion curve. 

The paper is organized as follows: we first describe the different EOS setups implemented in the experiment, then we discuss the resolution and calibration and finally we present the measurements of the FEL temporal profile as a function of input beam energy and discuss the unusual pulse splitting behavior of the waveguide FEL regime. 

\section{Electro-optic sampling of THz FEL radiation pulses}
The EOS technique utilizes the crystal birefringence induced by an applied electric field (the THz FEL pulse in this case) in an electro-optic crystal. The birefringence can be probed by measuring the induced phase retardation between the different polarization components of an infrared (IR) laser pulse passing through the crystal at the same time as the THz pulse. For the EO-active material ZnTe, cut in the (110) plane, the effect is maximized for a normally incident THz pulse orthogonally polarized with respect to the [001] axis of the crystal such that the index ellipsoid eigenvectors align at 45 degrees with respect to electric field direction. The induced birefringence can then most effectively be probed by a linearly polarized IR pulse with polarization either parallel or orthogonal to the incoming THz polarization \cite{faure2004modelling,chen2001electro}.

The difference in refractive index between the slow and fast axis of the crystal is proportional to the instantaneously applied THz electric field \cite{chen2001electro}, and  the resulting phase retardation $\Gamma$ can be written as \cite{van2006coherent}
\begin{equation}
    \Gamma= \frac{2\pi L n_0^3\, r_{41}}{\lambda}T_{\text{corr}} E_{\text{THz}} \equiv \frac{E_{\text{THz}}}{E_{\text{cal}}}
    \label{Eq:EO_cal}
\end{equation}
in which $L=0.2$~mm is the crystal length, $n_0=2.85$ is the refractive index of ZnTe at $\lambda$=800~nm and $r_{41}=4\times 10^{-12}$~m/V is the EO coefficient of ZnTe \cite{tilborgtemporal}. In general, the correction coefficient $T_{\text{corr}}$ includes the transmission losses through the crystal and the dephasing effects between IR and THz that are frequency dependent so that one has to be careful in interpreting the EOS signal as a direct field measurement. Nevertheless, in the spectral range relevant for this paper (<0.5\,THz), crystal losses (50\,\%) are frequency independent and dephasing is negligible \cite{tilborgtemporal},
so we can use the approximation  $T_{\text{corr}} \approx 0.5$. In practical units, Eq. \ref{Eq:EO_cal} implies that an applied field of $E_{\text{cal}}$=13.6 MV/m induces a 1 rad phase shift. 

Different EOS encoding setups can be implemented to measure the induced phase retardation and reconstruct the original THz field profile.
In the simplest case, one can use a polarizer after the crystal, perpendicularly oriented to the incoming polarization of the IR beam so that no light is transmitted in the absence of THz radiation. With an electric field present, the IR picks up the EO-induced phaseshift, allowing light to be transmitted. While signal to noise (STN) ratio is maximal in this cross-polarized scheme, as explained below, the drawback is that the measurement is insensitive to the sign of $\Gamma$ and the applied electric field.

In order to obtain sign information, a balanced detection scheme can be used. Here a quarter waveplate (QWP) is inserted after the ZnTe crystal with the fast axis oriented 45$^\circ$ with respect to the probe beam to create circular polarization. A Wollaston prism is then used after the EO crystal to split the two polarization components of the probe beam on the detector. The EO signal can then be measured looking at the difference between the IR intensities in the two polarizations. 

In both cases, the relation between detected intensity and induced phaseshift can be evaluated using Jones matrices yielding respectively
\begin{align}
    I^{\text{cross}}_\text{det}(\Gamma)&= \frac{I_0}{2}(1-\cos\Gamma)\approx \frac{I_0}{4}\,\Gamma^2 \label{Icross}\\
    I^{\text{bal}}_\text{det}(\Gamma)&=I_0\sin\Gamma\approx I_0\,\Gamma
    \label{Ibal}
\end{align}
in which $I_0$ is the input intensity of the probe laser. The approximations in \eqref{Icross} and \eqref{Ibal} hold for small phaseshifts $\Gamma\ll 1$. 

While at first look the balanced detection scheme appears to yield better signal for smaller electric fields (or induced phase shifts), one has to be careful to consider the equivalent noise in each case. In the cross-polarized scheme, the signal is detected on top of a background which is only limited by the camera electric noise and the residual drive laser fluctuations leaking through the finite extinction ratio of the polarizer pair. Assuming this latter component can be made dominant in the setup, the rms relative noise readout of 0.3 parts per million, so that the smallest detectable phase shift is $\sim$ 1.1\,mrad. 

The reference waveform in the balanced detection scheme is instead calculated as the difference between the two streaks without a signal present. The noise is significantly reduced by the subtraction as IR intensity fluctuations appear equally in both polarization components. Still, fluctuations in the laser spectrum, position and polarization add up to an rms noise corresponding to an effective phaseshift of 6~mrad. This value (notably significantly larger than the cross-polarized threshold) should then be considered as the minimum bound on a detectable single shot signal for the balanced detection scheme.

With an ultrashort IR probe pulse and a mechanical delay line, the temporal evolution of the applied field can be analyzed by scanning the delay between probe and THz pulse to directly retrieve $\Gamma$ as a function of time. This approach, referred to as multishot-EOS, is sensitive to shot to shot relative time-of-arrival jitter between the THz (and hence the e-beam) and the IR probe pulse which can be quite large in our setup (0.6 ps rms). 

\begin{figure}[h!]
\centering
\includegraphics[width=1\textwidth]{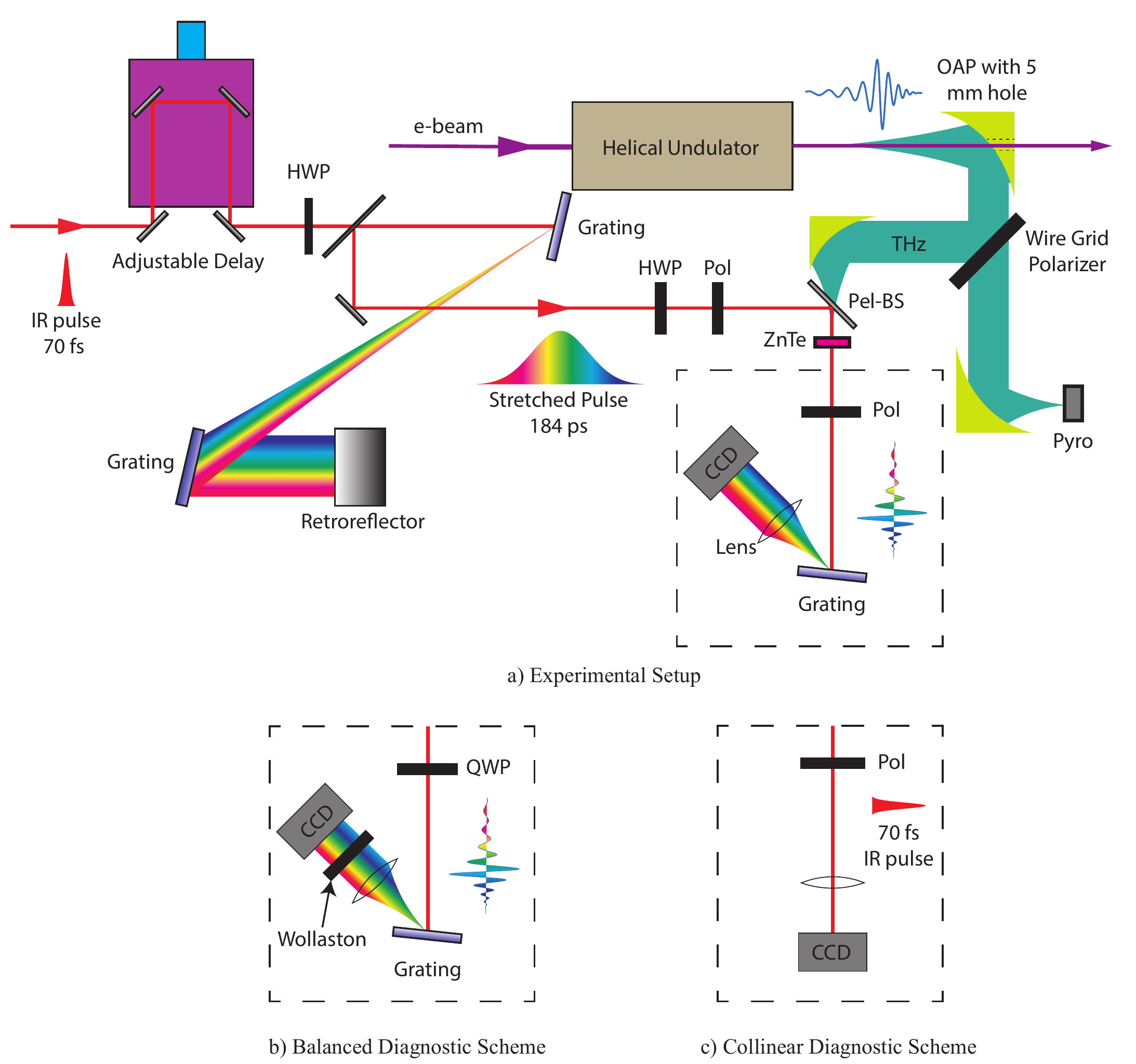}
\caption{In a), the experimental setup is portrayed for the cross-polarized single shot EOS implementation. HWP, Pol, Pel-BS stands for half-waveplates, polarizers and pellicle beamsplitter respectively. In b) we show the balanced detection setup with the addition of the Wollaston prism and the quarter waveplate (QWP). Finally, in c) we show the setup for the multishot spatial diagnostics where the grating stretcher is bypassed using a two redirecting mirrors to keep the IR pulse length short.}
\label{fig:Setup}
\end{figure}

To avoid the time resolution blurring due to this time-of-arrival jitter, a single shot scheme can also be implemented. The broadband IR pulse is stretched from 70~fs to hundreds of picoseconds using a pair of gratings \cite{treacy1969}  (see Fig.\,\ref{fig:Setup}). The spectro-temporal correlation of the IR pulse at the crystal causes then different wavelengths to experience different phase shifts. The temporal profile therefore can be reconstructed simply by using a grating, a lens, and a CCD to record the spectrum of the IR beam after the analyzer. 



The experiment has been carried out at the UCLA Pegasus photoinjector laboratory \cite{maxson,alesini} where an electron beam of up to 200 pC of charge is generated illuminating the photocathode in an S-band gun with a FWHM 100 fs UV pulse. The bunch is lengthened by longitudinal space charge forces and then it is recompressed to 1.5 ps rms at the entrance of the undulator by properly tuning the phase and amplitude of a booster linac cavity to achieve a total beam energy in the range  5-7\,MeV, (relativistic $\gamma$ factor\,=\,10-14). The helical undulator is composed of 28 full periods with period length of 32 mm, has an initial field amplitude 0.73\,T, and is parabolically tapered by adjusting the gap to maintain resonance with the decelerating electron beam. The 4.54~mm waveguide diameter yields a zero-slippage matching resonant frequency of 0.16~THz for $\gamma = 10.5$ electrons. At the time of these experiments, poor emittance and alignment in the undulator limited the charge transmission to $<$ 20 $\%$ of the injected charge which reduced the FEL output energy from the best performances by nearly an order of magnitude. 

At the undulator output, the THz radiation is collected with a 2" off-axis parabolic (OAP) mirror with a 5~mm diameter hole for the electron beam to pass through. A wire-grid polarizer (16 wires/mm) splits the circularly polarized THz radiation in horizontal and vertical components. The transmitted radiation is focused down on a reference pyro-electric detector while the reflected radiation is overlapped with the IR probe beam in the EO crystal. The IR is first sent through a delay stage to adjust arrival time over a span of 1~ns.

In the spectrally-encoded setup, we use a half-waveplate (HWP) before the stretcher to tune the IR polarization and maximize reflectivity off the gratings. The stretcher consists of two 1200 lines/mm gratings and a retroreflector to outcouple the stretched pulse after two passages on each of the gratings. Prior to redirecting the beam onto the ZnTe using a pellicle beamsplitter (Pel-BS, fully transmissive in the THz and 50\% reflective in the IR regime) we adjust its polarization back to maximize the EO effect. For the cross-polarized diagnostic scheme in Fig.\,\ref{fig:Setup}a, a second polarizer is placed after the ZnTe crystal, 90$^\circ$ orientated to the first (Cross Polarization). 

In Fig.\,\ref{fig:Setup}b, we show the balanced detection scheme where we replace the analyzer with a quarter-waveplate with the fast axis oriented 45$\dg$ to the IR polarization creating circularly polarized light. The Wollaston prism (25 mrad separation angle) is placed 8.5~cm before the camera and splits the two polarizations components so that they are captured in the same exposure on the CCD detector (2.2~mm apart) to be analyzed. 

In Fig.\,\ref{fig:Setup}c the multishot-EOS scheme is portrayed. Here, the IR stretcher is bypassed to maintain a short IR probe beam and data is acquired by scanning the delay line to measure the temporal profile of the THz pulse. This configuration can be used to measure the transverse spatial profile of the THz which can be effectively visualized after the analyzer by imaging the ZnTe crystal plane on to the CCD camera as shown in Fig.\,\ref{fig:Multishot}a. Multiple CCD images are saved at each delay line position and averaged in post-processing to smooth out the shot-to-shot fluctuations in the FEL energy due to the beam charge variations.

\section{Multishot EOS setup for FEL pulse characterization}

It is useful to start the analysis of the results from the measurements performed in this latter configuration. The data is recorded for input electron beam energy of $\gamma=11.2$ to maximize the FEL output. 

\begin{figure}[h!]
    \centering
\centering
\includegraphics[width=1\textwidth]{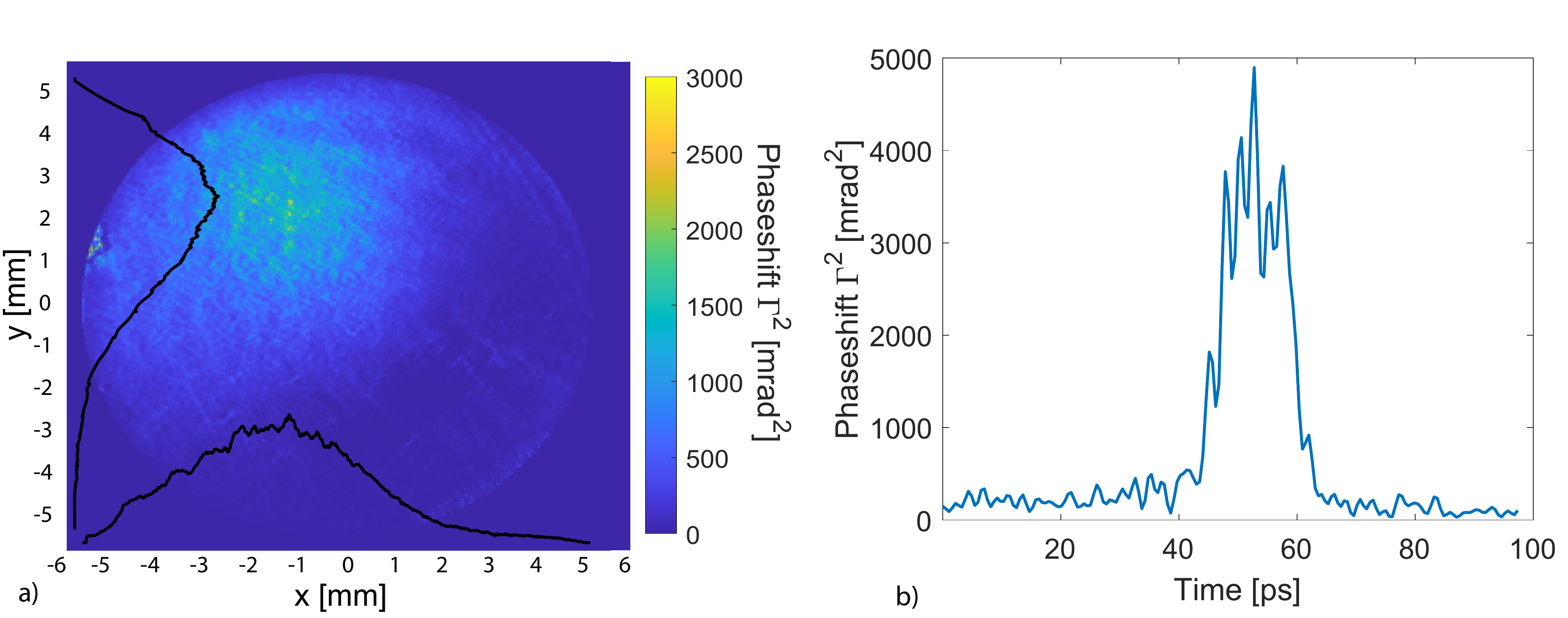}
	\caption{a) Transverse profile of THz FEL radiation pulse at the ZnTe crystal plane. The circular outline of the 10 mm diameter ZnTe crystal is visible and slightly clips the beam. b) Temporal profile of the THz radiation pulse obtained from scanning the delay line in the multishot measurement setup (16 fs/step).}
	\label{fig:Multishot}
\end{figure}

\begin{figure}[h!]
    \centering
        \begin{subfigure}{0.48\textwidth}
			\includegraphics[scale=0.4]{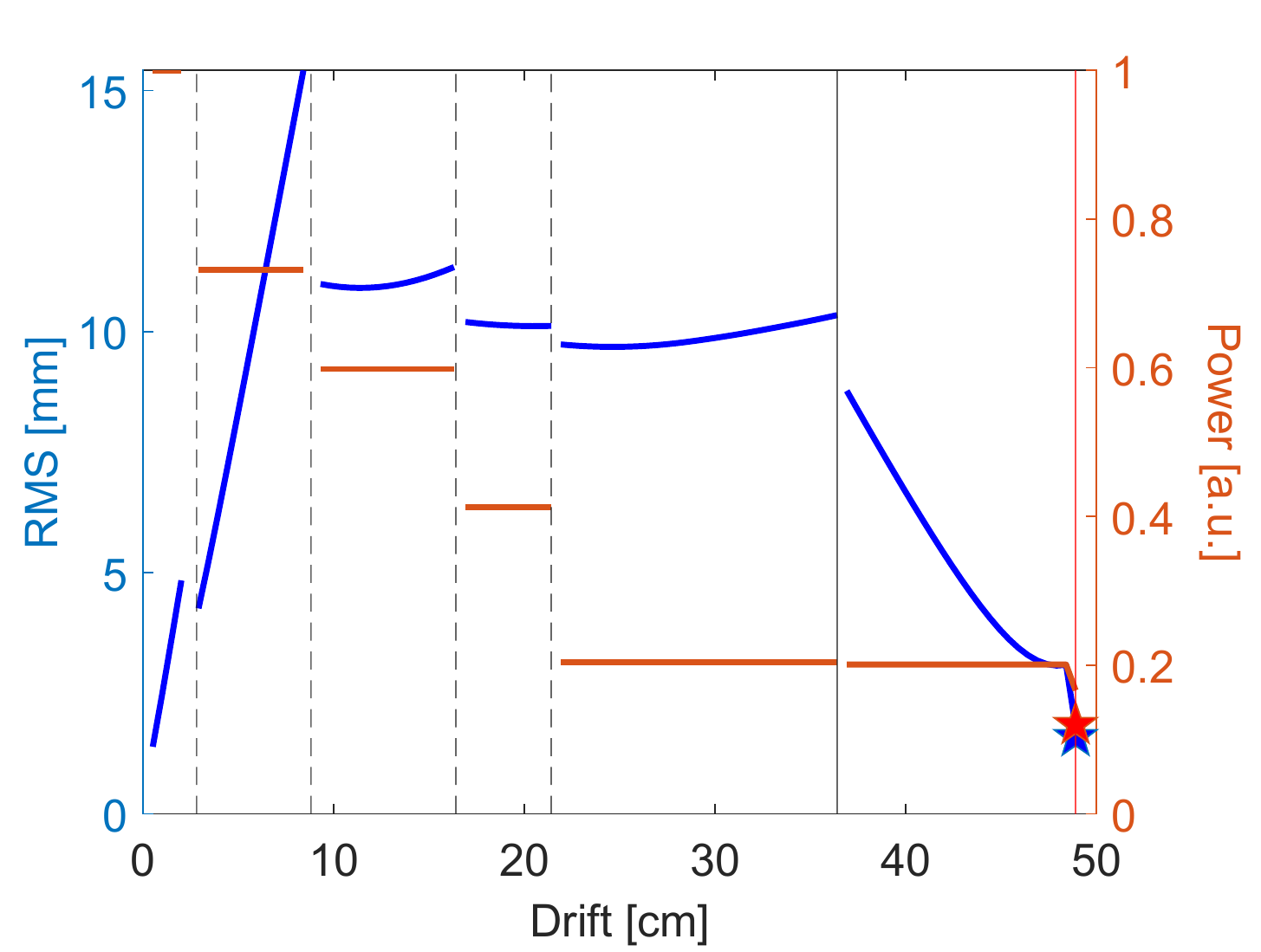}
	\end{subfigure}
	 \caption{Results of THz propagation simulations. The rms diameter of the intensity radiation profile and the relative power are plotted along the THz transport. The measured data points at the ZnTe plane are indicated as stars.}
\label{THz_Transport}
\end{figure}

The measured radiation rms spot size 1.65~mm favorably compares with the estimate obtained using an ABCD matrix-based radiation propagation code to model THz transport from the exit of the undulator waveguide to the ZnTe crystal plane shown in Fig.\,\ref{THz_Transport}. The initial field profile is assumed to be the ideal TE$_{11}$ mode of a circular waveguide. Due to the long wavelength, the aperture of the vacuum beamline and the optics severely affect the THz propagation. Taking into account the 50\,\% wire grid polarizer splitting factor (and assuming 70\,\% transmission for the vacuum window but no other losses in the propagation), the simulation predicts 87\% loss to the ZnTe crystal. To validate this model a 10~mW CW 135~GHz source was placed at the entrance of the undulator waveguide to measure transmittance through the line. Due to input coupling losses and absorption on the waveguide wall through the undulator, 2~mW were detected at the exit the undulator. Power-meter measurements in the reference path and in the EOS path yielded 0.15~mW, indicating higher than expected losses $>$90\% likely due to the limited transmission and reflection of our optics. 

After background subtraction, we can average over the strongest 30\,\% of the shots to take into account the effect of the charge transmission fluctuations and plot the EO signal as a function of time delay. This is shown in Fig.\,\ref{fig:Multishot}b from which we can estimate an rms THz pulse length of 5~ps.

We can estimate the energy contained in the THz pulse as
\begin{equation}
    U_\text{THz}=\epsilon_0 c A_\text{THz} E_\text{cal}^2 \sum \Gamma^2 \Delta t = 0.52~\mu\text{J}
\end{equation}
in which $A_{\text{THz}}$ the transverse area of the radiation and $E_{\text{cal}}=13.6~$MV/m our phaseshift to electric field calibration from Eq. \eqref{Eq:EO_cal}. 
Taking into account the measured transmission losses on the line we estimate the THz energy at the undulator exit to be 7.3 $\mu$J, significantly lower energies than in previous experiments \cite{TESSATRON}, but consistent with the lower charge transmission through the undulator.

The THz frequency oscillation is visible in the data, but note that the time-of-arrival jitter between electron and probe beam affects the temporal resolution in this multishot configuration. As discussed later in the paper, the rms jitter is measured to be 0.6~ps, therefore significantly blurring the oscillations in the THz field profile.

\section{Single-shot EOS measurements}
\begin{figure}[h!]
    \centering
    \includegraphics[scale=0.42]{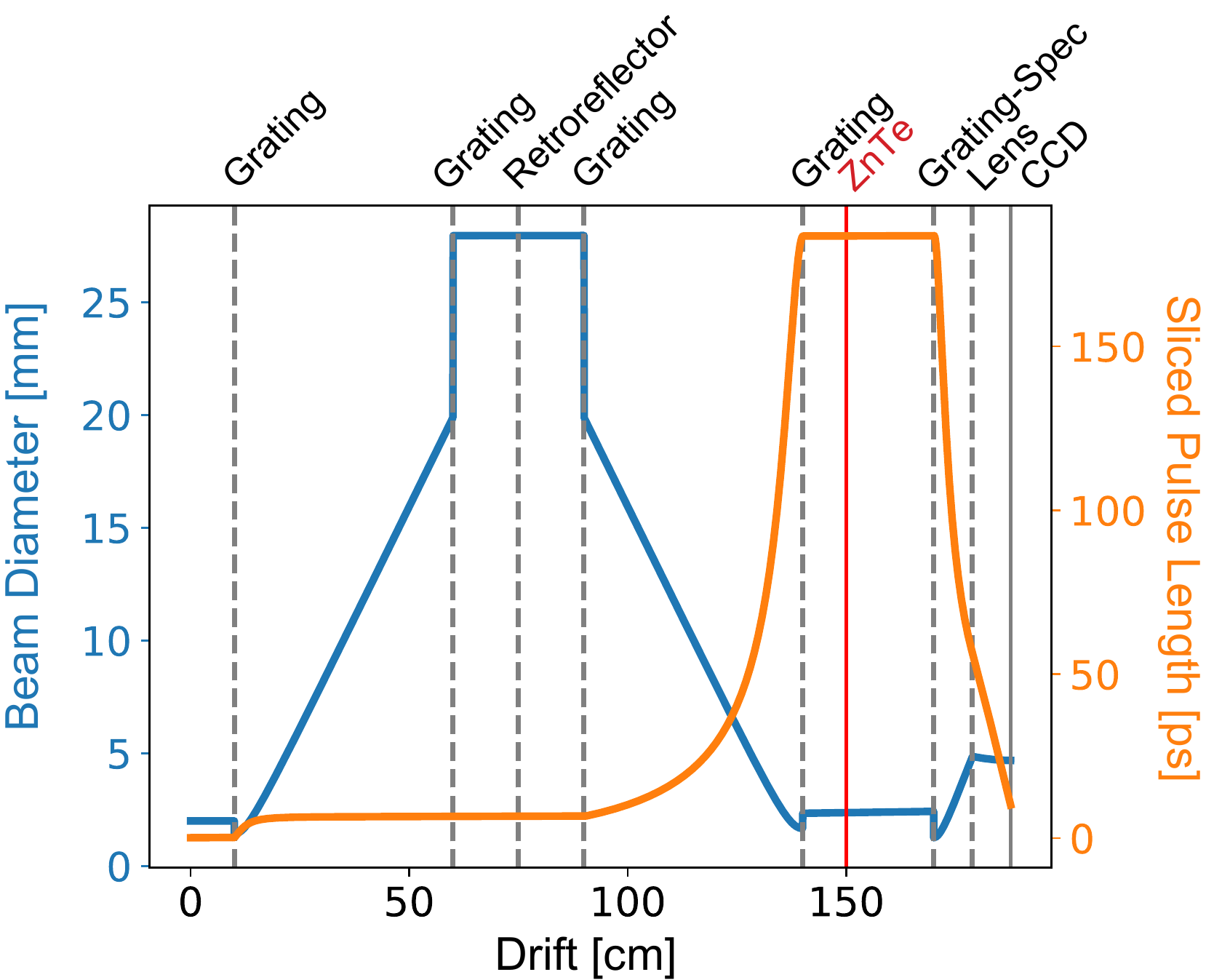}
    \caption{Probe IR beam propagation simulations. The 1/e$^2$ beam diameter (blue) and full pulse length (orange) are shown along the propagation line. The locations of the various optical elements are indicated by gray dotted lines.}
    \label{fig:Kostenbauder}
\end{figure}

In order to allow single-shot EOS measurements, we modify the setup as shown in Fig.\,\ref{fig:Setup}a. The IR probe beam transport was modeled with $4 \times 4$ Kostenbauder matrices, a formalism which computes the spatio-temporal evolution of Gaussian pulses through the optical system allowing for pulse length and spotsize estimates as shown in  Fig.\,\ref{fig:Kostenbauder}.

Simple formulas can also be used to understand quantitatively the transport. Group-delay-dispersion after the grating pair can be written as
\begin{equation}
    \text{GDD}=-\frac{\lambda^3}{\pi c^2 d^2} \frac{L_\text{sep}}{\cos^2\theta_\text{out}}=-2.74\text{ps}^2
\end{equation}
with $\lambda$ = 800~nm being the center wavelength of our probe beam, $d$=1/1200~mm the line separation on the grating, $L_\text{sep}=$ 0.5~m the distance between the gratings and $\theta_\text{out}$ = 0.8~rad the first order diffraction angle \cite{trebino_book}. The resulting FWHM pulse length can be estimated as $\Delta t_\text{stretch}= \frac{\sqrt{\Delta t^4+16 \ln^2(2)\, \text{GDD}^2}}{\Delta t}=108.5\,\text{ps}$ assuming an initial transform limited pulse with FWHM $\Delta t = 70$\,fs. This value can be used to compute the 1/e$^2$ full pulse width (184 ps) which is in excellent agreement with the Kostenbauder calculation shown in Fig.\,\ref{fig:Kostenbauder}. Since the chirp is mostly linear over the full spectral bandwidth, the spectro-temporal correlation is therefore 184 ps / 23 nm = 8.0~ps/nm. We also note that the full pulse width at the ZnTe plane in this scheme sets the length of the active single-shot measurement window. For longer THz pulses it will be necessary to take more than one shot at different positions of the IR delay line.

The final spectrometer consists of a 1200~lines/mm grating where the IR is incident at a slight angle of 104 mrad with a lens ($f$=88~mm) placed at a focal length distance behind the grating to collimate the diffracted light on to a CCD sensor and visualize the IR spectrum. The dispersion from the grating is $dx/d\lambda=2.04e5$ yielding a spectral calibration of 0.018 nm/px for the 3.75 $\mu$m CCD pixel size. Putting all together, we expect a time-calibration of 8.0~ps/nm $\times$ 0.018~nm/px or 0.15~ps/px.

\begin{figure}[h!]
    \centering
		\includegraphics[scale=0.8]{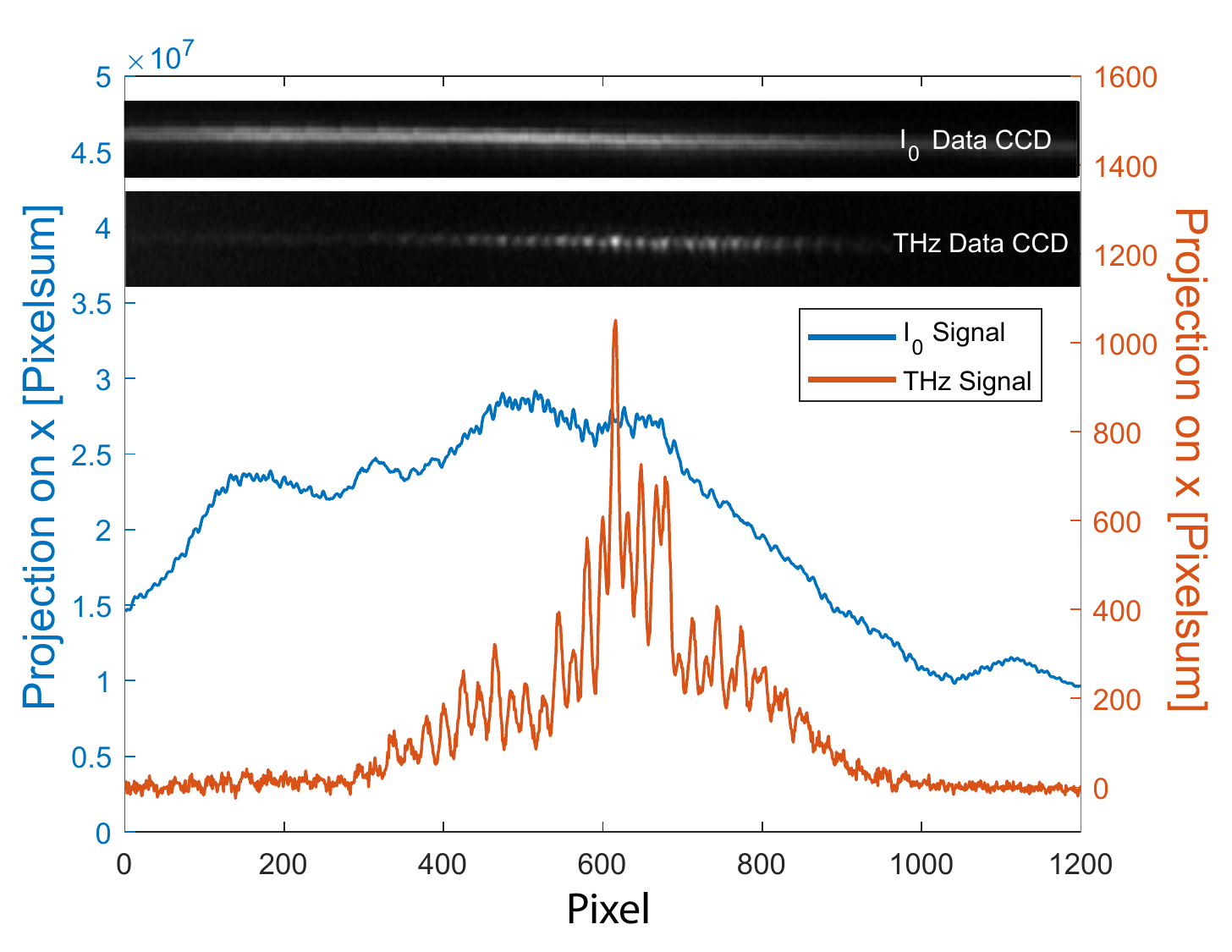}
	\caption{Raw data from EOS CCD for the reference spectrum and sample trace taken in the cross-polarized configuration with the THz FEL on. The x-axis projections from the data are presented in blue and orange respectively.}
    \label{fig:I0}
\end{figure}

Sample data acquired in this configuration is shown in Fig.\,\ref{fig:I0} where the modulation imprinted by the THz field oscillation can be clearly seen on the IR spectrum. Note that due to a slight asymmetry in the polarizer, the temporal profile shows alternating lesser and stronger peaks in Fig.\,\ref{fig:I0} which we interpret as the positive and negative peaks of the THz field illuminating the crystal. 

\begin{figure}[h!]
    \centering
\includegraphics[width=1\textwidth]{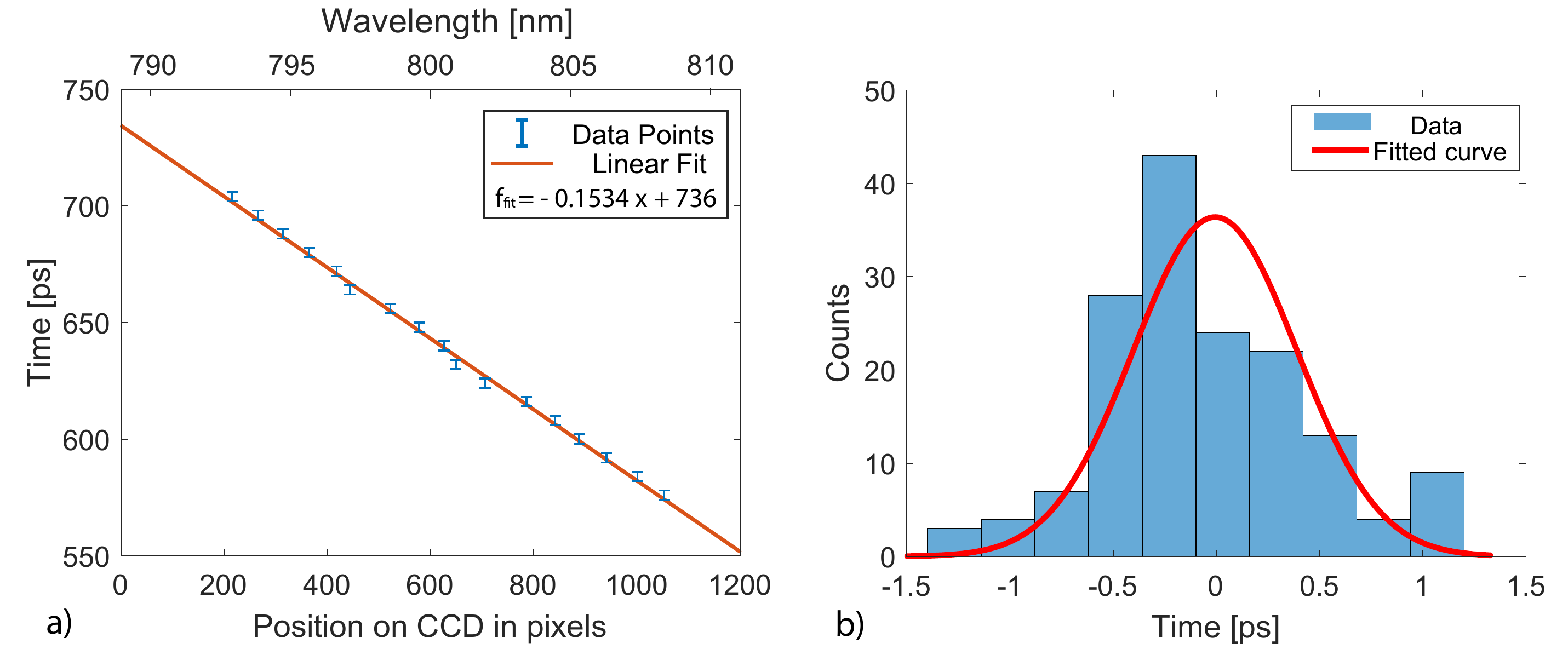}
\caption{a) Peak position in THz waveform as a function of delay line. The wavelength scale on the spectrometer CCD is shown on the top axis b) Histogram of the time of arrival of the THz pulse, as measured by recording the peak position of the single shot EOS waveforms. The redline shows a gaussian fit with 0.5\,ps rms.}
	\label{fig:Time}
\end{figure}

After background subtraction, the signal is projected on to the spectrometer axis and then normalized by the original spectrum of the IR pulse $I_0$ (also shown) to retrieve the induced phaseshift. In the presence of a THz signal, we can verify the predicted temporal calibration by adjusting the IR delay line which shifts the signal horizontally on the CCD. The shift is measured through the change in the peak position and compared to the change in optical delay time. The data is presented in Fig.~\ref{fig:Time}a. The linear fit gives a slope of 0.153~ps/px in excellent agreement with previous Kostenbauder-based calculations. Rescaling the horizontal axis in Fig.\,\ref{fig:I0} using the measured calibration directly yields a temporal profile of the THz pulse.

It is interesting to note that for a fixed delay line position, the shape of the EOS trace is essentially constant for each shot only slightly shifting horizontally on the CCD.  In single-pass high gain FELs starting from noise one expects the waveform to randomly change between the shots \cite{Bonifacio1994}, but in the UCLA THz FEL case, the amplification process is seeded by the sub-wavelength compressed electron bunch and the field is naturally phase-locked with the beam. The horizontal jitter in the traces is instead due to the fluctuations in relative time-of-arrival between the IR and the e-beam induced by amplitude and phase jitter in the RF gun and linac. 

In order to quantify this jitter, the signal peak position is acquired for 200 shots at a fixed delay-stage position and plotted as a histogram in Fig.\,\ref{fig:Time}b. An analysis of the histogram data yields an rms time-of-arrival jitter between the IR pulse and the THz FEL pulse of 0.6~ps, consistent with previous measurements at Pegasus \cite{scoby2010}. 

We also note here that due to the finite temporal resolution of the diagnostics, the signal does not go to zero in between the peaks. In practice, the temporal resolution of the technique has various contributions, including the velocity mismatch of the THz pulse and the IR in the ZnTe crystal as well as the finite pulse length of the unstretched IR pulse. The largest contribution though is due to the resolution of the spectrometer which can be estimated by analyzing the y-axis projection on the CCD sensor which has a $1/e^2$-width of 4 pixels. This translates into a blurring in the time profile of the THz field of 0.6 ps. In our analysis we assume a gaussian point spread function and in post-processing the data, we use deconvolution in Fourier space in order to take this effect into account in the single-shot time traces. In addition, a low-pass filter with cutoff at 0.5~THz is also applied in Fourier space to smooth out the signals.

\section{Pulse splitting in waveguide THz FEL}

\begin{figure}[h!]
    \centering
	\begin{subfigure}{0.48\textwidth}
    	\includegraphics[scale=0.38]{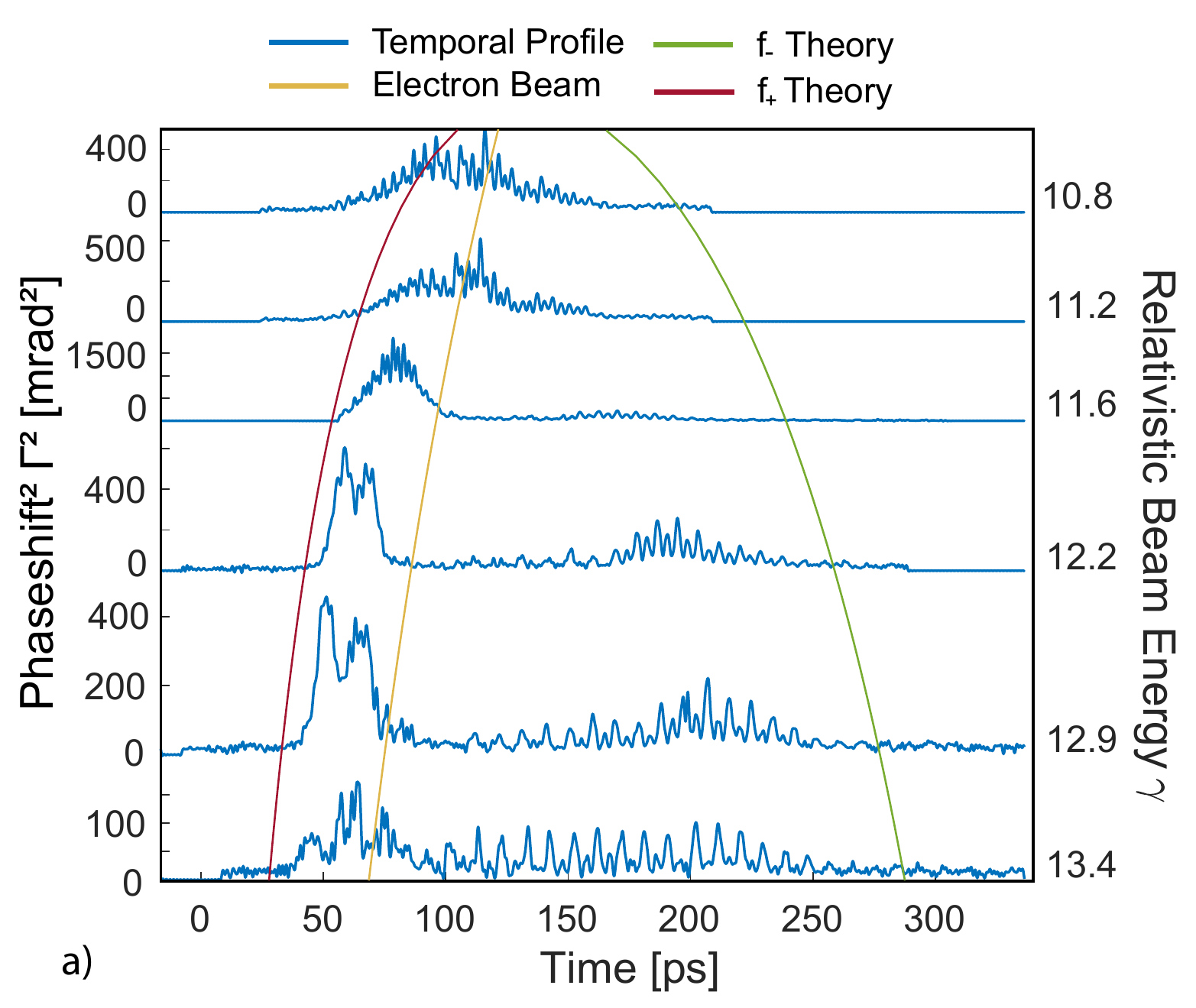}
	\end{subfigure}
    \begin{subfigure}{0.48\textwidth}
    \includegraphics[scale=0.44]{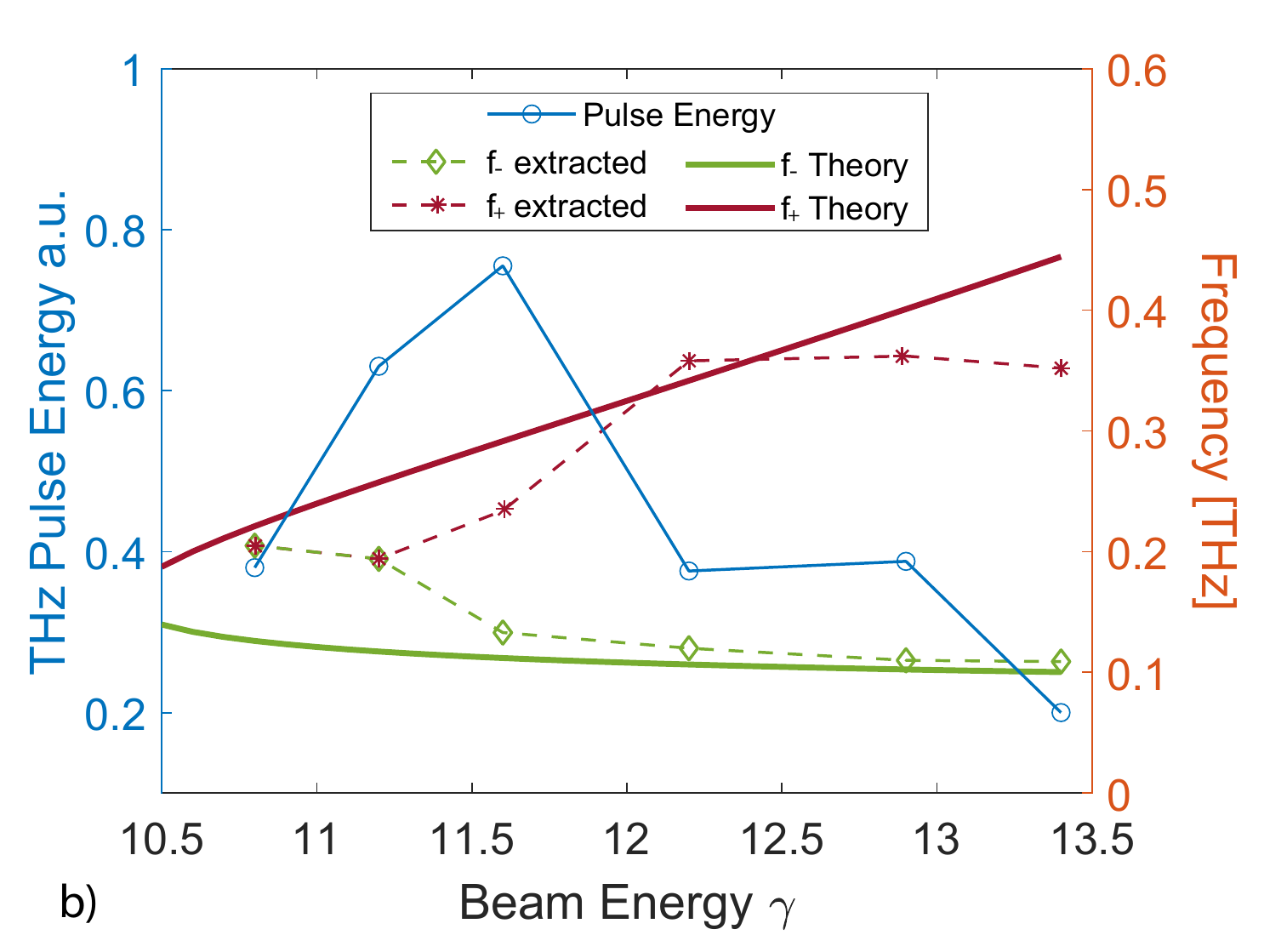}
    \end{subfigure}
	\caption{a) Temporal profiles for different beam energies. The colored lines represent the theoretical arrival times for the e-beam as well as low ($f_-$) and high ($f_+$) resonant frequencies b) Pulse energies and peak frequencies extracted from the data as a function of beam energy. For $\gamma>11.6$ we plot two frequencies for the trailing and leading pulse respectively. The analytical solutions for the low and high frequency branches are also plotted.}
	\label{fig:SingleShot}
\end{figure}

Single-shot THz FEL radiation waveforms are then acquired as a function of the beam energy at the entrance of the undulator and are presented in Fig.\,\ref{fig:SingleShot}a. The linac phase is adjusted to vary the beam energy and the bunch length into the undulator. 

As the input beam energy approaches the zero-slippage resonant condition, the waveforms compress in time and become more intense, consistent with the finding in \cite{TESSATRON}. Conversely, as the beam energy is increased, a characteristic temporal structure develops. For $\gamma > 11.6$ the pulse is observed to separate into a high and low frequency component with the high frequency component arriving first, followed by a delayed low frequency pulse. 

By integrating the signal (proportional to the square of the electric field) over time we recover the energy in the radiation pulse (shown in Fig.\,\ref{fig:SingleShot}b). It is not surprising that the strongest signal is measured for $\gamma=11.6$, slightly off the zero-slippage resonant condition. This is partly due to finite emittance and energy spread, but also higher e-beam energies significantly improve transmission through the undulator, increasing the effective charge participating to the interaction.

In the interferometry measurements originally reported in \cite{TESSATRON}, pulse splitting could not be directly observed, but is consistent with the picture that for an input energy off the zero slippage condition, there are two phase-velocity resonant frequencies in the system with positive and negative group velocities with respect to the e-beam. An analytical approximation of the phase-resonant frequencies for relativistic beams is $f_\pm=f_{zs} \left( 1+\frac{2\Delta \gamma}{\gamma_0}\right)\left( 1 \pm \sqrt{\frac{2\Delta\gamma}{\gamma_0}}\right)$ where $\Delta \gamma = \gamma - \gamma_0$ and $\gamma_0$ is the zero-slippage resonant energy \cite{TESSATRON}.

In the cross-polarized scheme the Fourier transform can not be used to faithfully retrieve the spectral content of the radiation pulse because the sign of the phaseshift is not encoded in the temporal profile. Nevertheless, the dominant frequency components in the different pulses can be extracted by averaging the distance between the peaks. Assuming consecutive peaks represent positive and negative electric fields, the frequency is calculated as the inverse of the mean separation between alternating peaks. Fig.\,\ref{fig:SingleShot}b shows the dominant frequency components in the leading and trailing edge of the pulse. In comparing these with the analytical prediction lines (also shown), it should be noted that the latter are obtained from calculating the intersection between the beam phase velocity and the waveguide dispersion curve, but finite beam emittance and energy spread effects are known to broaden and shift the spectral response. 


Using the retrieved frequency it is possible to better understand the time traces for the different input beam energies. The yellow line in the plot shows the time-of-arrival of the e-beam at the exit of the undulator which accounts for the difference in e-beam velocity. We then have to consider the different propagation times for the THz radiation in the $L_{wg}= 1$~m long metal waveguide of radius $R$ = 2.27~mm which is used to meet the zero-slippage condition. Different THz frequencies travel at different velocities, resulting in different times of transit. Using the expression for the group velocity of the lowest frequency TE$_{11}$ mode (which has the strongest coupling with the e-beam current),  $v_g(f)=\sqrt{1-\frac{c k_\bot}{2\pi f}^2}$ where $k_\bot=1.8412/R$ is the transverse wavenumber, we can calculate the additional contribution as $\Delta t= L_{wg}/v_g(f)$.

Accounting for frequency dependent group velocities results in a significant difference in arrival times between the two resonant frequencies. The two solid lines in Fig.\,\ref{fig:SingleShot}a represent the theoretical estimates obtained by summing up the e-beam velocity contribution and the frequency dependent group velocity of the THz pulse through the undulator-waveguide. The data clearly shows that in the beam rest-frame the two THz phase-velocity resonant frequencies have a positive and negative group velocity respectively.

\section{Balanced Detection}

In order to retrieve a sign-resolved temporal profile of the radiation pulse and the full radiation spectrum we then implemented the balanced detection scheme adding the quarter-wave plate and the Wollaston prism (setup shown in Fig.\,\ref{fig:Setup}c) to visualize both polarization components at the same time on the CCD camera. 

\begin{figure}[h!]
    \centering
	\begin{subfigure}{0.48\textwidth}	
    	\includegraphics[scale=0.42]{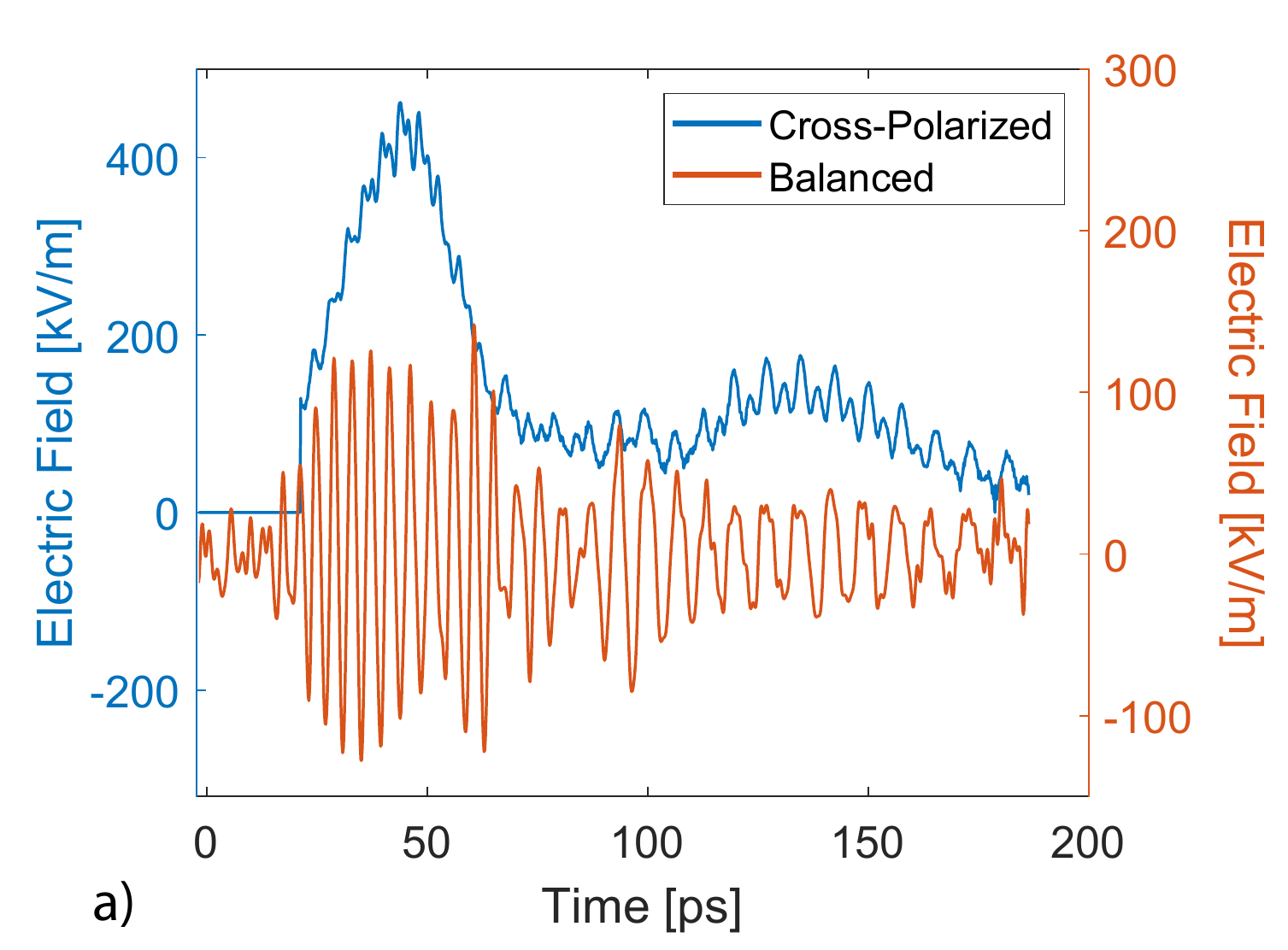}
	\end{subfigure}
    \begin{subfigure}{0.48\textwidth}
			\includegraphics[scale=0.42]{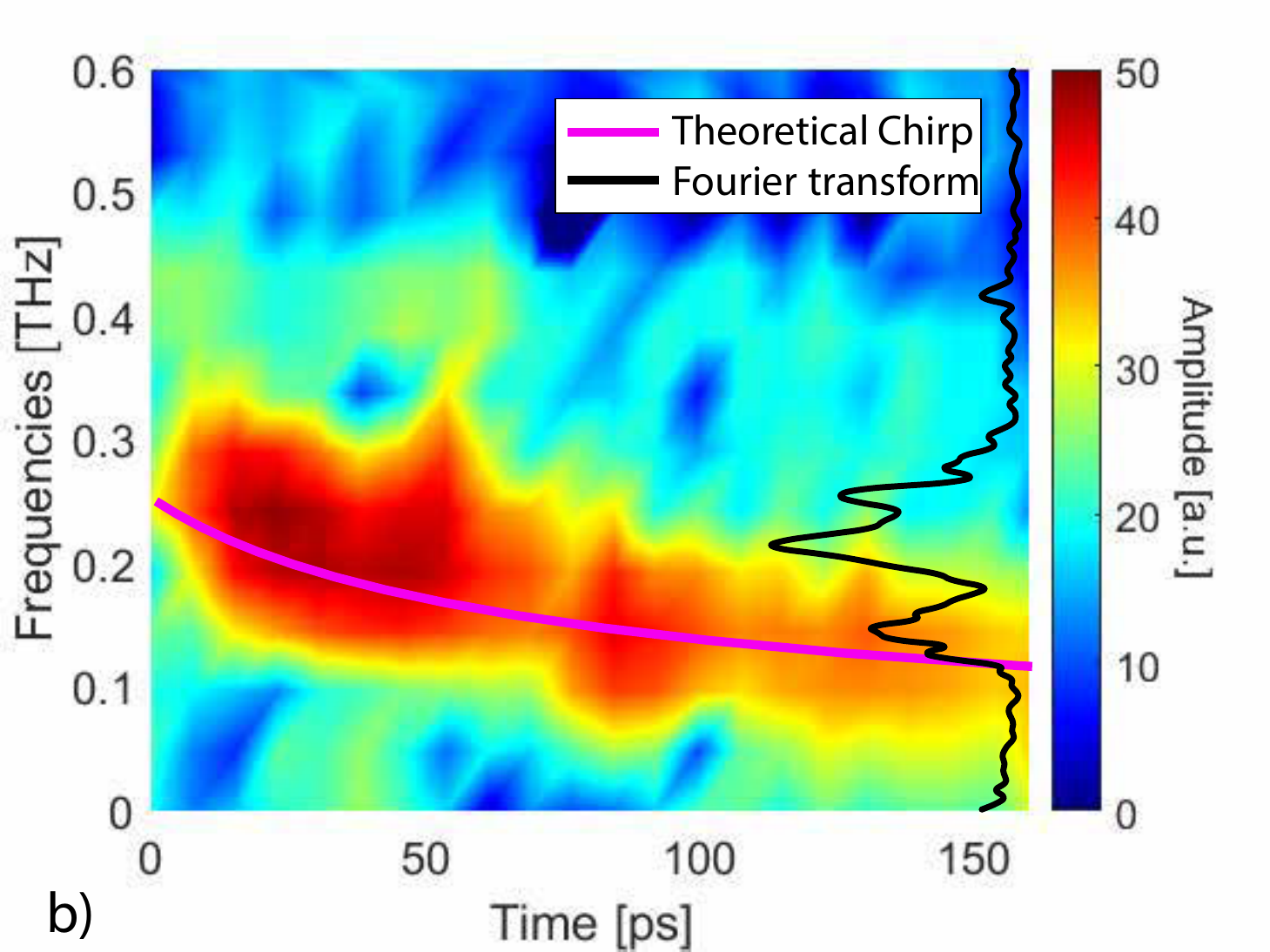}
	\end{subfigure}
	\caption{a) Sign-resolved electro-optic sampling trace of the THz FEL pulse from the spectrally-encoded balanced detection scheme at $\gamma=11.6$ compared to the cross-polarized data. b) Spectrogram plot obtained using a short-time window Fourier transform. The full Fourier transform is projected on the right. The group delay dispersion curve is also shown and is found to be consistent with the pulse chirp.}
	\label{fig:balanced}
\end{figure}

After background removal, the induced phase shift is then retrieved by subtracting and normalizing the two streaks on the CCD. The electron beam was set to $\gamma=11.6$ for these measurements, but the transmitted beam charge through the undulator was less than 40 pC, greatly reducing the FEL output. In addition, as previously discussed, the signal-to-noise ratio in this scheme is significantly lower, since the induced phase shift is a small perturbation on top of an intense spectrum trace. Averaging over multiple shots can be used to reduce the noise and results in the sign-resolved temporal profile portrayed in Fig.\,\ref{fig:balanced}a. The temporal calibration is the same as in the cross-polarized setup as no optics before the ZnTe crystal were changed. A clear frequency chirp along the pulse can be observed. 

We use a short-time window Fourier transform algorithm on the THz waveform to visualize this time-frequency correlation in the pulse as a 2D spectrogram (contour plot in Fig.\,\ref{fig:balanced}b). The observed frequency chirp compares well with the expected delay induced by the frequency dependent group velocity (group dispersion delay curve, plotted in purple). In addition, the full radiation pulse spectrum (black curve on the right) is calculated as the Fourier transfom of the entire time-trace and extends from 0.1 to 0.25~THz confirming the broad-band nature of the waveguide FEL interaction.

\section{Conclusion}

In summary, we have presented the results of the first EOS-based time-domain characterization of a tapered undulator THz-FEL operating in the zero-slippage regime. Pulse splitting in the THz waveforms is observed for input beam energies higher than the zero-slippage condition in agreement with theoretical expectations. 

Using a balanced detection scheme, we also clearly visualize a strong temporal chirp in the radiation pulse. In principle, since the pulse shape is fixed for a given input electron beam, it could be possible to add negative dispersion (for example using chirped mirrors as in \cite{strecker:dispersioncompensation}), reverse the waveguide-induced frequency dependent-delay, recompress the FEL pulse, and achieve higher THz peak intensities. 

Exploiting the large spectral bandwidth of the FEL pulse, the EOS setup that we demonstrated here could also be used for single-shot time-domain spectroscopy both in static and pump-probe modalities. The EOS diagnostics could also be applied to effectively measure THz waveforms from other electron-beam-based long-pulse THz radiation schemes such as Cherenkov wakefield radiation from dielectric and corrugated structures \cite{oshea2016, bane:corrugatedpipeTHz}. 

\section*{Acknowledgements}
This work was supported by NSF grant PHY-1734215 and DOE grant No. DE-SC0009914. 
\section*{Disclosure}
The authors declare no conflicts of interest.
\section*{Data Availability}
Data underlying the results presented in this paper are not publicly available at this time but may be obtained from the authors upon reasonable request.

\bibliography{EOS_THzFEL}

\end{document}